\documentclass[fleqn,usenatbib]{mnras}

\usepackage[T1]{fontenc}
\usepackage{ae,aecompl}

\usepackage{graphicx}	% Including figure files
\usepackage{amsmath}	% Advanced maths commands
\usepackage{amssymb}	% Extra maths symbols
\usepackage{multicol}
\usepackage{bm}
\usepackage{booktabs}

\title[Morphologically-Selected Mergers at $\lowercase{z}\sim 2$]{The MOSDEF Survey: Differences in SFR and Metallicity for Morphologically-Selected Mergers at $\lowercase{z}\sim 2$$^{1}$}

\author[K. Horstman et al.]{Katelyn Horstman,$^{2}$\thanks{E-mail: katelynhorstman@gmail.com}
Alice E. Shapley,$^{2}$
 Ryan L. Sanders,$^{3,4}$ \newauthor
 Bahram Mobasher,$^{5}$
 Naveen A. Reddy,$^{5}$
 Mariska Kriek,$^{6}$ \newauthor
 Alison L. Coil,$^{7}$
 Brian Siana,$^{5}$
 Irene Shivaei,$^{8,4}$
 William R. Freeman,$^{5}$ \newauthor
 Mojegan Azadi,$^{9}$
 Sedona H. Price,$^{10}$
 Gene C. K. Leung,$^{7}$
 Tara Fetherolf,$^{5}$ \newauthor
 Laura de Groot,$^{11}$
 Tom Zick,$^{6}$
 Francesca M. Fornasini,$^{9}$
 Guillermo Barro$^{12}$
\\
$^{1}$Based on data obtained at the W.M. Keck Observatory, which is operated as a scientific partnership among the California Institute of \\ Technology, the University of California,  and the National Aeronautics and Space Administration, and was made possible by the generous  \\ financial support  of the W.M. Keck Foundation.\\
$^{2}$Department of Physics and Astronomy, University of California, Los Angeles, 430 Portola Plaza, Los Angeles, CA 90095, USA\\
$^{3}$Department of Physics, University of California, Davis, 1 Shields Avenue, Davis, CA 95616, USA\\
$^{4}$Hubble Fellow\\
$^{5}$Department of Physics and Astronomy, University of California, Riverside, 900 University Avenue, Riverside, CA 92521, USA\\
$^{6}$Astronomy Department, University of California at Berkeley, Berkeley, CA 94720, USA\\
$^{7}$Center for Astrophysics and Space Sciences, Department of Physics, University of California, San Diego, 9500 Gilman Drive, \\ La Jolla, CA 92093, USA\\
$^{8}$Steward Observatory, University of Arizona, 933 N Cherry Ave, Tucson, AZ 85721, USA\\
$^{9}$Harvard-Smithsonian Center for Astrophysics, 60 Garden Street, Cambridge, MA, 02138, USA \\
$^{10}$Max-Planck-Institut f\"ur Extraterrestrische Physik, Postfach 1312, Garching, 85741, Germany \\
$^{11}$Department of Physics, The College of Wooster, 1189 Beall Avenue, Wooster, OH 44691, USA\\
$^{12}$Department of Phyics, University of the Pacific, 3601 Pacific Ave, Stockton, CA 95211, USA
}

\begin{document}
\label{firstpage}
\maketitle

\begin{abstract}

    We study the properties of 55 morphologically-identified merging galaxy systems at $z\sim 2$. These systems are flagged as mergers based on features such as tidal tails, double nuclei, and asymmetry. Our sample is drawn from the MOSFIRE Deep Evolution Field (MOSDEF) survey, along with a control sample of isolated galaxies at the same redshift. We consider the relationships between stellar mass, star formation rate (SFR), and gas-phase metallicity for both merging and non-merging systems.  In the local universe, merging systems are characterized by an elevated SFR and depressed metallicity compared to isolated systems at a given mass. Our results indicate SFR enhancement and metallicity deficit for merging systems relative to non-merging systems for a fixed stellar mass at $z \sim 2$, though larger samples are required to establish these preliminary results with higher statistical significance. In future work, it will be important to establish if the enhanced SFR and depressed metallicity in high-redshift mergers deviate from the ``fundamental metallicity relation," as is observed in mergers in the local universe, and therefore shed light on gas flows during galaxy interactions. 
\end{abstract}


\section{Introduction} 
\label{sec:intro}

Galaxies grow through merging events and the smooth accretion of baryons and dark matter. Mergers reflect the hierarchical growth of structure formation within the $\Lambda$CDM cosmological framework. The rate at which dark matter halos merge as a function of mass, mass ratio, and cosmic time is well predicted by cosmological simulations of structure formation \citep[e.g.,][]{cole2008,fakhouri2010}. At the same time, actual observations of the frequency of galaxy mergers over a wide range in redshift is of key importance in constraining the nature of galaxy assembly. In addition, assessing the effect of mergers on the properties of interacting systems provides key constraints on the flow of gas and the formation of stars during these important stages of galaxy evolution.

In the local universe, the Sloan Digital Sky Survey (SDSS) has yielded statistical samples of galaxy pairs at $z \sim 0$, pre-coalescence \citep{ellison2008, scudder2012, scudder2015, patton2011, patton2013}. These systems are identified by a projected radius separation of between 30 to 80 kpc and a radial velocity difference between 200 and 500 km s$^{-1}$. Merging systems selected using the above criteria are characterized by enhanced star-formation rate (SFR) of $\sim 60$\% out to 30 kpc and depressed gas-phase metallicity of $\sim 0.02-0.05$~dex \citep{scudder2012,ellison2008} relative to isolated systems of the same stellar mass. Predictions from simulations by groups such as \cite{hopkins2008} and \cite{bustamante2018} are consistent with observations in the local universe, explaining how inflow of gas into the central regions of the merging galaxies both increases SFR and lowers the gas phase oxygen abundance of the ISM.
SFR enhancement has also been detected in merging pairs out to $z \sim 1$ \citep{lin2007,wong2011}. 

Merging systems have now been identified out to  $z \sim 6$ \citep{ventou2017}. At $z > 1$, mergers are commonly flagged either 
through galaxy pairs or by observing morphological features indicative of disturbance. A frequent method for identifying merging systems is through photometric pairs. Galaxies within a small projected radius that have small differences in photometric redshift have been studied to assess merger fraction and merger rate in the early universe \citep{williams2011, man2012, man2016, mantha2018,silva2018}.
Other studies have used rest-frame ultraviolet spectra to identify mergers spectroscopically \citep{tasca2014, ventou2017}. However, rest-frame UV features are sensitive to large-scale galaxy outflows \citep{pettini2001, shapley2003, steidel2010}, limiting the accuracy with which merger dynamics can be measured. 
Another common technique for flagging mergers is to use morphological features to recognize coalescing systems. Classifiers use visual identifiers such as tidal tails and bridges, and double nuclei \citep{lofthouse2017} to categorize merging systems. Additionally, galaxies are identified as interacting based on non-parametric morphological statistics \citep{lotz004, con2014,cibinel2019}.

A recent study conducted using the MOSFIRE Deep Evolution Field (MOSDEF) survey \citep{kriek2015} focuses on determining SFR and gas-phase metallicity at a given stellar mass to compare merging and isolated systems at $z \sim 2$ \citep{wilson2019}. Spectroscopic redshifts are used to identify merging systems, with the corollary that the sample is sensitive to early-stage, pre-coalescence mergers, since at least two, spatially-distinct, emission-line redshifts must be measured to define a merging system. In the current work, we use a complementary approach to identify mergers at $z\sim 2$, based on morphological classifications in the Cosmic Assembly Near-infrared Dark Energy Legacy Survey (CANDELS) morphology catalog \citep{kartaltepe2015}. This method may select mergers over a larger range of interaction stages, including later-stage, coalesced systems \citep{cibinel2019}.  We then trace key scaling relations among galaxy properties such as stellar mass, SFR, and metallicity for both mergers and isolated galaxies, comparing the two samples. 

In Section~\ref{sec:observations_samples}, we present the details of the MOSDEF survey and our merger selection criteria. Section~\ref{sec:results} investigates the relationship between SFR and metallicity for a given stellar mass at $z \sim 2.3$ for both merging and non-merging systems. In Section~\ref{sec:discussion}, we conclude by presenting a discussion of our results and describing future work. Throughout this paper, we adopt cosmological parameters of $H_0= 70$ km s$^{-1}$ $\mathrm{Mpc}^{-1}$, $\Omega_M = 0.30$, and $\Omega_{\Lambda}=0.70$.

\section{Observations and Samples}
\label{sec:observations_samples}

\subsection{The MOSDEF Survey}
\label{sec:MOSDEF}

\begin{figure}
\begin{centering}
\includegraphics[scale=0.5]{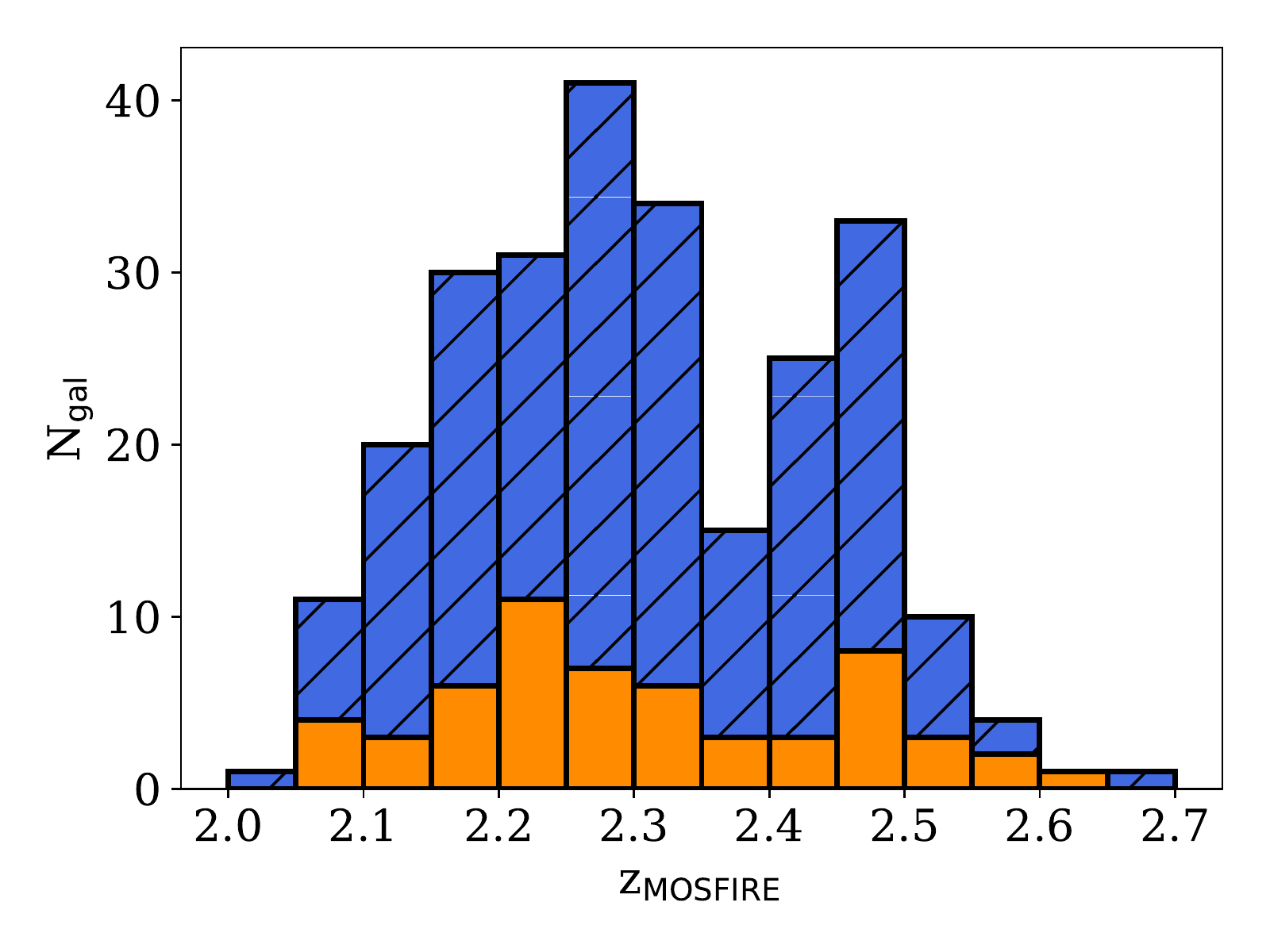}
\caption{Redshift distributions. The MOSDEF parent sample is shown in hatched blue, while the merger sample is shown in solid orange. The median redshift of the parent sample is $z_{\rm med}=2.29$. The median redshift of the merger sample is $z_{\rm med}= 2.28$.}
\label{fig:redshift_histogram}
\end{centering}
\end{figure}

\begin{figure*}
\includegraphics[scale=0.92]{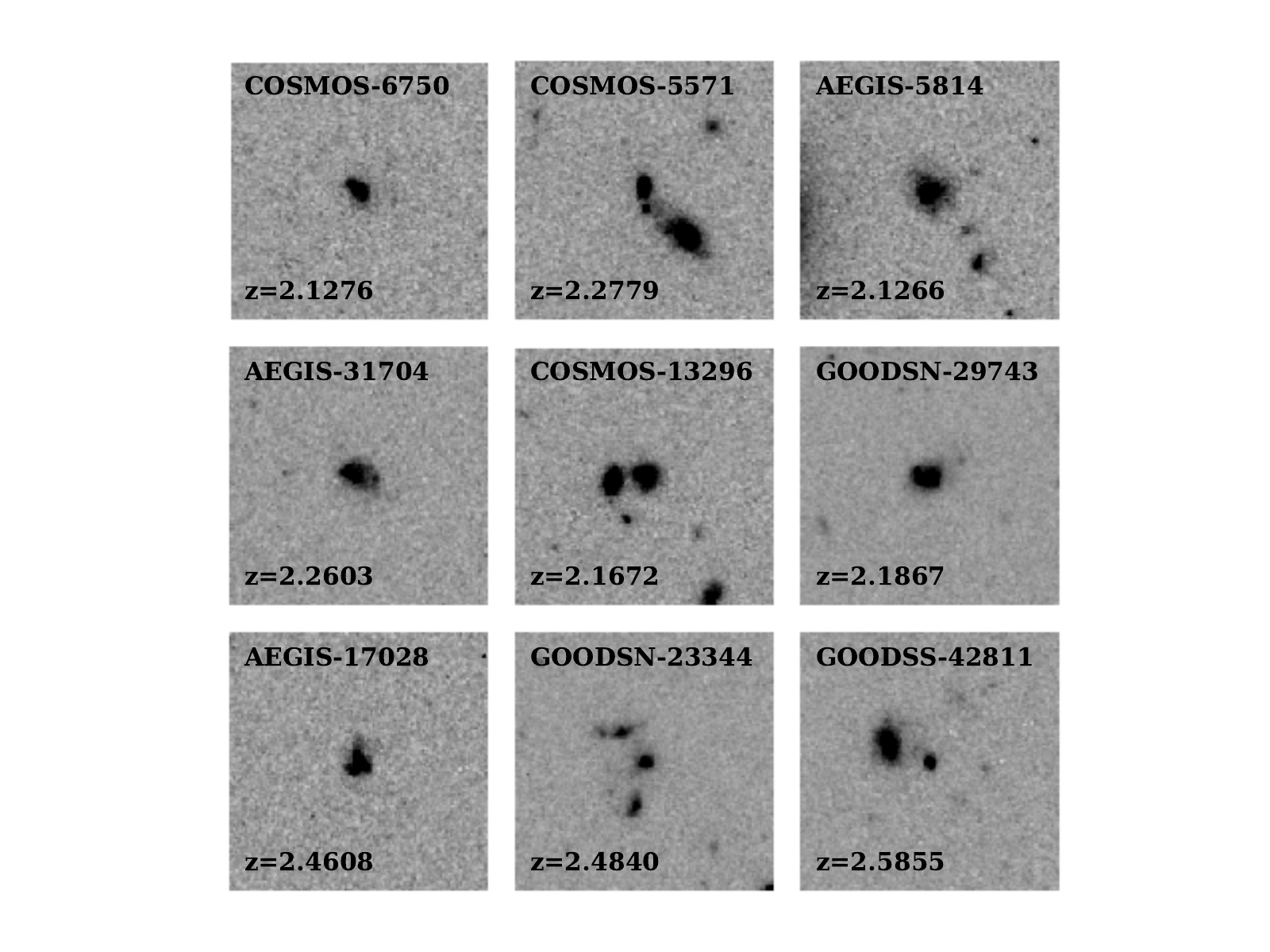}
\caption{Merging galaxies at $2.0\leq z \leq2.7$. Each panel represents a distinct morphologically-classified merger (class 3 or 4, i.e., high-confidence classification as a merger or interaction, as described in the text), labeled with its field in the upper left corner, and its v4.1 ID and redshift in the bottom left corner. Each panel is 10" by 10". Postage stamps were made using CANDELS imaging data processed by the 3D-HST survey \citep{skelton2015}.}
\label{fig:merger_stamps}
\end{figure*}

\begin{figure*}
\includegraphics[scale=0.92]{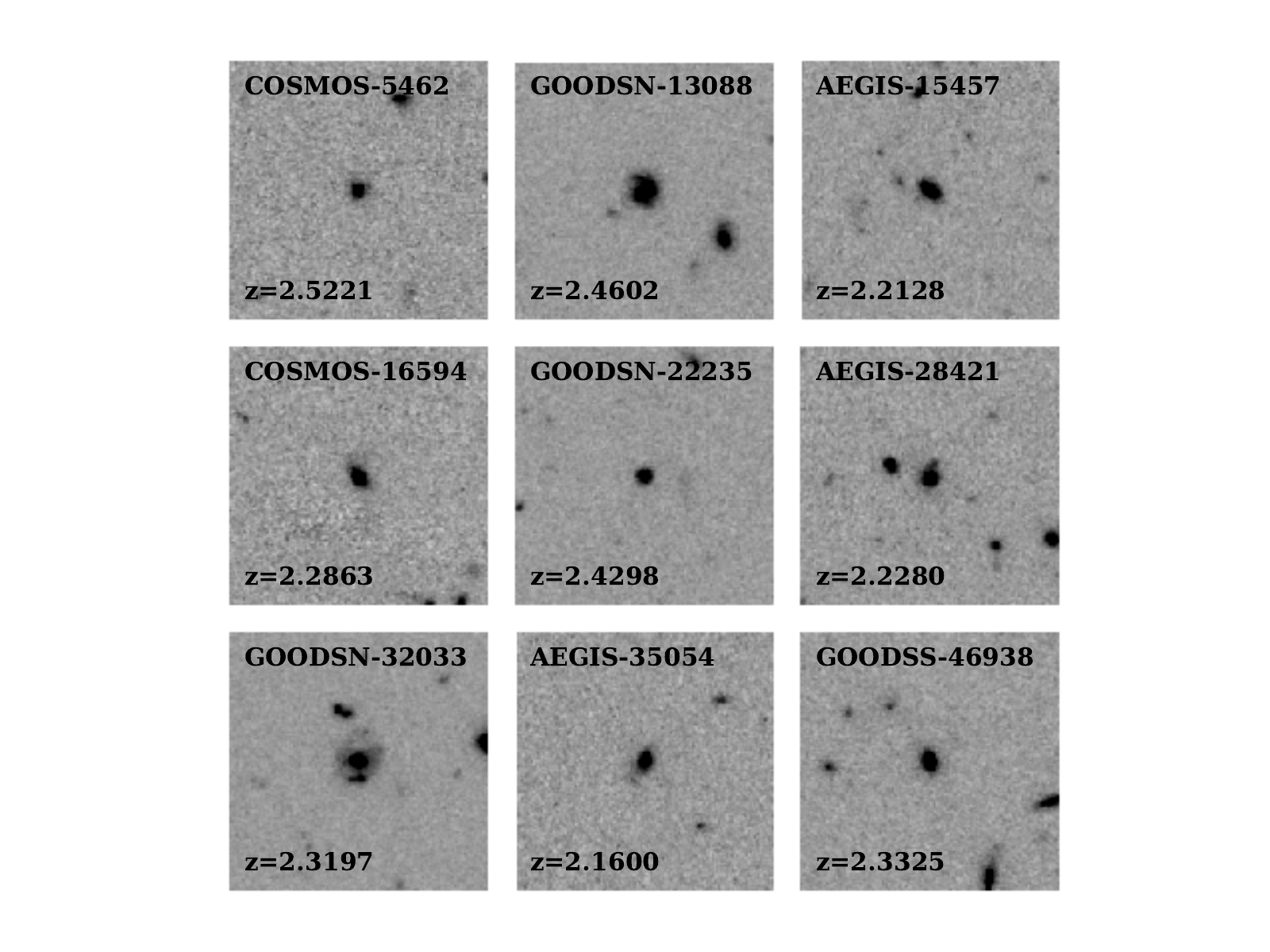}
\caption{Isolated galaxies at $2.0\leq z \leq2.7$. All labels and scaling as in Figure 2.}
\label{fig:non_merger_stamps}
\end{figure*}

Our sample of high-redshift galaxies is drawn from the MOSDEF survey. The MOSDEF survey used the Multi-Object Spectrometer for Infra-Red Exploration \citep[MOSFIRE;][]{mclean2012} for 48.5 nights during 2012--2016 to acquire rest-frame optical spectra of $\sim 1500$ galaxies at a redshift between $1.4\leq z \leq3.8$. Galaxies were targeted in three redshift intervals where the strongest emission lines are found within near-infrared windows of atmospheric transmission: $1.37\leq z \leq1.70$, $2.09\leq z \leq2.61$, and $2.95\leq z \leq3.80$. Based on spectroscopic and photometric catalogs compiled as part of the 3D-HST survey \citep{brammer2012, momcheva2016, skelton2015}, we selected galaxies located in the following CANDELS fields: AEGIS, COSMOS, GOODS-N, GOODS-S, and UDS \citep{grogin2011, koekemoer2011}. All target galaxies have extensive multi-wavelength photometric coverage \citep{skelton2015} used to derive stellar population parameters and photometric redshifts. A full description of the survey and data reduction is presented in \citet{kriek2015}. 

Using MOSFIRE emission-line fluxes and existing imaging data, 
we derived multiple galaxy properties. 
H$\alpha$ SFRs (SFR(H$\alpha$)) were obtained from dust-corrected and slit-loss-corrected H$\alpha$ luminosities based on the calibration of \citet{hao2011} for a \citet{chabrier2003} initial mass function (IMF), and assuming the \citet{cardelli1989} extinction law. 
Stellar masses (M$_*$) were found from emission-line corrected photometry using the fitting program FAST \citep{kriek2009}, assuming delayed-exponential star-formation histories of the form SFR$\propto t \exp(-t/\tau)$, solar metallicity, a \citet{chabrier2003} IMF, and the \citet{calzetti2000} dust attenuation curve.
The best-fit stellar population model to the galaxy
spectral energy distribution (SED) yielded an independent estimate of the SFR, SFR(SED).
To estimate oxygen abundances, we used the N2=log([NII]$\lambda$6584/H$\alpha$) and O3N2=log(([OIII]$\lambda$5007/H$\beta$)/([NII]$\lambda$6584/H$\alpha$)) calibrations from \citet{pettini2004}. The following 
calibrations are for N2 and O3N2 respectively:

\begin{equation}
    12+ \mathrm{log(O/H)_{N2}}=8.90 +0.57\times \mathrm{log(N2)}
\end{equation}

\begin{equation}
    12+ \mathrm{log(O/H)_{O3N2}}=8.73 -0.32\times \mathrm{log(O3N2)}
\end{equation}

As in \citet{sanders2018}, we constrained our parent MOSDEF sample by requiring: a redshift at $2.0\leq z \leq2.7$, $\log(M_*/M_{\odot}) \geq9.0$, and a detection of H$\alpha$ and H$\beta$ at a signal-to-noise ratio S/N$\geq$3 and free from significant sky-line contamination. Additionally, we excluded objects identified as AGN based on X-ray emission, {\it Spitzer}/IRAC colors, or [NII]/H$\alpha$
ratios $\geq 0.5$  \citep{coil2015,azadi2017,leung2019}. In total, 250 galaxies were selected. Figure~\ref{fig:redshift_histogram} shows the redshift distribution of our sample. 

\subsection{Merger Selection}
\label{sec:mergers}

To identify mergers within the parent MOSDEF sample, we made use of the CANDELS morphology catalog of Karteltepe et al. (private communication, 2019).
In this catalog, structural features such as tidal tails, double nuclei, and asymmetry were used to flag mergers and interactions. Three to five human classifiers assigned a confidence class 1--4 to each candidate merging system to indicate the robustness of its merger classification. To obtain the cleanest
sample of morphologically-classified mergers, we restrict our ``merger" sample to the two highest-confidence classes, 3 and 4. Class 4 mergers are galaxies that are unanimously classified as a merger or interaction by all classifiers. Class 3 mergers are galaxies where at least 66\% of classifiers agreed the galaxy was a merger or interaction \citep{kartaltepe2015}. Class 1 and 2 mergers were removed from both the parent sample and the ``merger" sample entirely to avoid ambiguity in classifying galaxies definitively as mergers or non-mergers.

From the 250 galaxies in our $z\sim 2$ MOSDEF parent sample, 
55 are identified as class 3 or 4 mergers, 16 as class 1 or 2 (and therefore removed), and 179 as non-merger controls. The redshift distribution of the merger sample is overplotted on the redshift distribution of the parent sample in Figure~\ref{fig:redshift_histogram}. The merger sample is characterized by a median redshift of $z_{\rm med}=2.28$, which is well-matched to that of the MOSDEF non-merger control sample ($z_{\rm med}=2.30$). Furthermore, a Kolmogorov-Smirnov (K-S) test of the two redshift distributions yields a p-value=0.40, indicating that the null hypothesis that the two distributions are drawn from the same parent distribution cannot be ruled out.  Figures~\ref{fig:merger_stamps}  and \ref{fig:non_merger_stamps} show representative examples of merging and non-merging galaxies in the MOSDEF sample, classified on the basis of morphology. In Figure~\ref{fig:merger_stamps}, we note features characteristic of merging galaxies, such as multiple nuclei within close proximity, as well as low-surface brightness tidal features and asymmetry. Non-merger galaxies in Figure~\ref{fig:non_merger_stamps} depict isolated nuclei and appear relatively smooth in shape compared to merging galaxies.  We note that stellar mass, SFR, and metallicity are estimated for morphological mergers using the same techniques as for isolated galaxies. Accordingly, for a subset of our merging sample we may be summing over multiple distinct merging components. However, the angular resolution of the MOSFIRE spectroscopic measurements used for estimating metallicities and SFRs do not allow for a more detailed decomposition. Furthermore, if multi-component stellar populations were a significant factor for the merging sample, we might expect poor stellar population model fits to the photometry when fitting with a single-component model, as reflected in the model reduced $\chi^2$ values. However, there is no evidence from
the reduced $\chi^2$ values for the stellar population fits for mergers and non-mergers that the merging systems have systematically worse fits.

High-redshift mergers were also identified in MOSDEF galaxies using spectroscopy and searching for multiple emission lines measures within the same spectroscopic slit and within a velocity separation of 500~km~s$^{-1}$ \citep{wilson2019}. However, in our analysis, merging galaxies were classified strictly using the CANDELS morphology catalog \citep{kartaltepe2015}.
We confirmed these classifications by detailed inspection of the {\it HST} WFC3/F160W images for the entire sample. 
Some of the visual criteria used for assigning class 3 or 4 ``merger" status may be associated with a merger that has already coalesced and would therefore not be flagged on the basis of two distinct emission-line redshifts, as in \citet{wilson2019}. Accordingly, our merger selection criteria may be sensitive to later-stage mergers than found in the sample of \citet{wilson2019}. We find evidence for this potential difference in the average merger stage probed given that $\sim50$\% of the morphologically-classified mergers only show signs of structural disturbance, as opposed to clear evidence for double or multiple nuclei or close companions. As such, our morphologically-classified merger sample is sensitive to mergers both pre-coalescence, when a distinct pair or multiple can be detected, and post-coalescence, when signs of morphological disturbance such as tidal tails and irregular outer isophotes
are still visible. At the same time, chance
projections are potentially a problem for our morphologically-based merger sample, which does not explicitly take into account
redshift information for each individual morphological subcomponent. Such chance projections would tend to dilute the measurement of underlying differences between the star-forming or chemical properties of mergers and non-merging systems, as they would not represent true merging systems. We note that there is only a small overlap of 4 galaxies between our morphological mergers sample and the merging pairs sample described in \citet{wilson2019}.

\section{Results}
\label{sec:results}

\begin{figure*}
\begin{centering}
\includegraphics[scale=0.53]{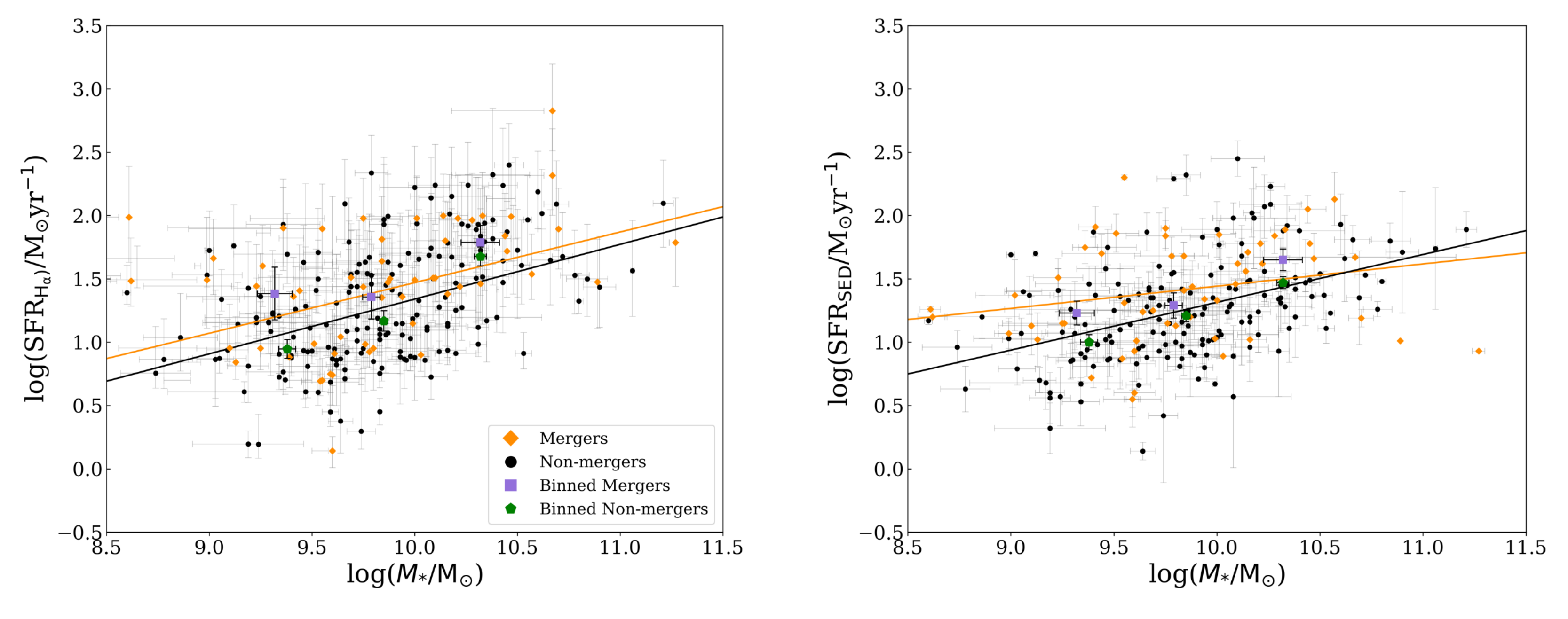}
\end{centering}

\caption{SFR-$M_*$ relation for mergers and non-mergers at $2.0 \leq z \leq 2.7$. \textbf{Left:} SFR(H$\alpha$) is estimated from the dust-corrected H$\alpha$ luminosity. Merging systems are indicated with orange diamonds while non-merging systems are indicated with black circles. A Bayesian linear regression was performed on both merging and isolated systems, shown with orange and black lines, respectively. Also shown are median SFR(H$\alpha$) values in three equal-sized bins in $M_*$, for both mergers (purple squares) and non-mergers (green pentagons).
\textbf{Right:} SFR(SED) is estimated from best-fit stellar population model to the broadband SED. Colors and symbols for both individual and stacked measurements are as in the left-hand panel. Similarly, a Bayesian linear regression was performed on both the merging and isolated systems, and the median SFR(SED) was calculated in three equal-sized bins in $M_*$ for both mergers and non-mergers.
}
\label{fig:SFRplots}
\end{figure*}

In the local universe, there are measurable differences in the SFRs and gas-phase oxygen abundances of merging systems 
as compared to isolated galaxies of similar mass. Analyzing the distinction between oxygen abundances in merging versus non-merging galaxies, \cite{ellison2008} found that for a given stellar mass, the mass-metallicity relation (MZR) corresponds to lower metallicities by up to 0.05 dex. Beyond the local universe, an enhancement in SFR in merging galaxies relative to isolated systems has been detected out to $z \sim 1$  \citep{wong2011}. To further understand these relationships at high redshift, we continue to explore stellar mass, metallicity, and SFR for merging and non-merging galaxies at $z\sim 2$ as in \citet{wilson2019}. 

\subsection{The SFR-$M_*$ Relation}
\label{sec:sfrm}

Here we explore the relationship between stellar mass and SFR for isolated galaxies and morphologically-classified mergers at $2.0\leq z \leq2.7$. 
For this analysis, we used two independent estimates of SFR: one derived from dust-corrected H$\alpha$ and H$\beta$ emission lines (SFR(H$\alpha$)) and another from stellar population synthesis model fits to broadband SEDs  (SFR(SED)). We show the correlations between SFR and $M_*$ in Figure~\ref{fig:SFRplots} for both merging and non-merging galaxies. The figure displays SFR(H$\alpha$) and SFR(SED), respectively, in the left and right panels. 

We performed separate linear regression fits to our merging and non-merging samples to quantify their respective correlations between SFR and $M_*$. For this analysis, we used the Bayesian linear regression algorithm described in \citet{kelly2007}, which takes into account errors in both x and y, and is implemented in the Python routine, LINMIX. This routine yields both the best-fit slope and intercept and their uncertainties. For the main regression analysis, we restricted the range of stellar masses to $9.0 \leq \log(\mbox{M}_*/\mbox{M}_{\odot})\leq 11.5$ and required $-0.5 \leq \log(\mbox{SFR/M}_{\odot}\mbox{yr}^{-1}) \leq 3.5$. 
For the SFR(H$\alpha$)-$\mbox{M}_*$ relation we find:

\begin{multline}
    \mathrm{log(SFR(H\alpha))_{\mbox{merger}}}= (-2.52 \pm 1.22) + \\
    (0.400 \pm 0.125) \times \mathrm{log(M_*/M_{\odot})_{\mbox{merger}}}
\end{multline}

\begin{multline}
    \mathrm{log(SFR(H\alpha))_{\mbox{non-merger}}}= (-2.98 \pm 0.68) + \\
    (0.432 \pm 0.069) \times \mathrm{log(M_*/M_{\odot})_{\mbox{non-merger}}}
\end{multline}
for merging and non-merging galaxies.
The slopes for the merger and non-merger relations are the same within their uncertainties, while there is a notable (though not significant) difference between the intercepts. The left panel of Figure~\ref{fig:SFRplots} is therefore suggestive of an elevated SFR(H$\alpha$) associated with merging galaxies at high redshift.

By comparing equations (3) and (4), we estimate the average difference between
the best-fit lines for merging and non-merging galaxies over the interval $9.0 \leq \log(M_*/\mbox{M}_{\odot}) \leq 11.5$ to be 0.13 dex. We also estimated the median SFR(H$\alpha$) values for merging and non-merging galaxy samples, given that the samples are very well matched in median stellar mass (i.e., $\log(M_*/\mbox{M}_{\odot})_{\rm med}=9.80$ for the merging sample and $\log(M_*/\mbox{M}_{\odot})_{\rm med}=9.84$ for the non-merging sample) and a K-S test of the two stellar mass distributions indicates a p-value=0.90, i.e., that the null hypothesis that two stellar mass distributions are drawn from the same parent population cannot be ruled out. We find that the median SFR(H$\alpha$) for the merging sample is $\log(\mbox{SFR}(\mbox{H}\alpha))_{\rm med}=1.48\pm0.04$, while the median for
the non-merging sample is $\log(\mbox{SFR}(\mbox{H}\alpha))_{\rm med}=1.26\pm0.06$. Accordingly, the median SFR(H$\alpha$) of the mergers is elevated by $\Delta\log(\mbox{SFR}(\mbox{H}\alpha))=0.22\pm 0.07$, with respect to that of the non-mergers. Figure~\ref{fig:SFRplots} (left) also shows the median SFR(H$\alpha$)
in three equal-size bins of stellar mass, for both mergers and non-mergers. We find that the average
offset among these more finely-sampled bins is $\langle\Delta\log(\mbox{SFR}(\mbox{H}\alpha))\rangle=0.24\pm 0.11$.
\\

For SED-based SFRs, we find:

\begin{multline}
    \mathrm{log(SFR(SED))_{merger}}= (-0.312 \pm 0.990) + \\
    (0.175 \pm 0.100) \times \mathrm{log(M_*/M_{\odot})_{merger}}
\end{multline}

\begin{multline}
    \mathrm{log(SFR(SED))_{non-merger}}= (-2.46 \pm 0.57) + \\
    (0.377 \pm 0.058) \times \mathrm{log(M_*/M_{\odot})_{non-merger}} 
\end{multline}
for merging and non-merging galaxies respectively.
The combined slope and intercept values for both merging and non-merging systems in the right panel of Figure~\ref{fig:SFRplots} do not indicate a clear elevation in SFR(SED) at fixed stellar mass at $z\sim 2$ in merging galaxies compared to non-merging galaxies. The offset between the median SFR(SED) values of the full merging and non-merging samples, while positive, is also not significant. We find a median of $\log(\mbox{SFR(SED)})_{\rm med}=1.34\pm0.12$ for the mergers, and $\log(\mbox{SFR(SED)})_{\rm med}=1.24\pm0.03$ for the non-mergers, corresponding to $\Delta \log(\mbox{SFR(SED)})=0.10\pm 0.12$. Figure~\ref{fig:SFRplots} (right) also shows the median SFR(SED)
in three equal-size bins of stellar mass, for both mergers and non-mergers. We find that the average
offset among these more finely-sampled bins is $\Delta\log(\mbox{SFR}(\mbox{H}\alpha))=0.16\pm 0.0.06$.

Using different methods to determine SFR may affect our ability to discern systematic differences between merging and non-merging systems. The H$\alpha$ SFR indicator is more sensitive to to short lived O-stars, while the SED indicator is sensitive to O-stars as well as longer lived B and A stars. SFR(H$\alpha$) accordingly tracks variations in SFR on shorter timescales, less than 100 Myr, while SFR(SED) is smoothed over longer timescales \citep{emami2019}. If high-redshift merging systems show elevated SFRs over timescales $\leq 100$~Myr \citep{fensch2017}, SFR(H$\alpha$) may be better suited to tracking such differences.

\subsection{The MZR Relation}
\label{sec:MZR}
\begin{figure*}
\includegraphics[scale=0.53]{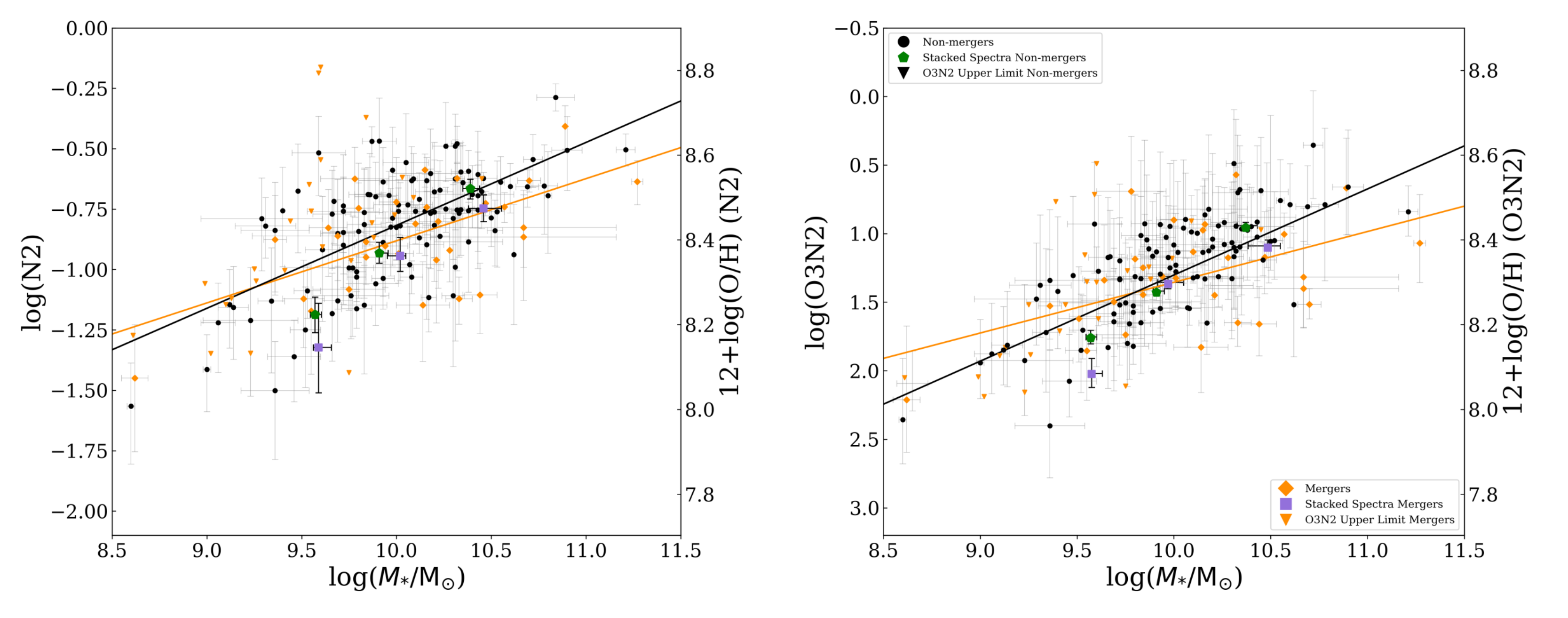}
\caption{MZR relation for mergers and non-mergers at $2.0 \leq z \leq 2.7$. \textbf{Left:} 12 + log(O/H) is determined from the N2 indicator and calibration of \citet{pettini2004}. Merging systems are indicated with orange diamonds while non-merging systems are indicated with black circles. Metallicity upper limits are indicated as orange and black triangles for mergers and non-mergers respectively. A Bayesian linear regression was performed on both merging and isolated systems shown with orange and black lines respectively. 
Additionally, we constructed median composite spectra in three bins of stellar mass for both merging and non-merging systems. Metallicities from the stacked spectra are shown with purple squares for merging systems and green pentagons for non-merging systems. The stacked measurements are consistent with the linear fit and suggest a dilution in metallicity at fixed stellar mass for merging systems compared to isolated systems. \textbf{Right:} 12 + log(O/H) is determined from the O3N2 indicator and the calibration of \citet{pettini2004}. Colors and symbols are as in the left-hand panel. Similarly, a  Bayesian linear regression was performed on both the merging and isolated systems. Metallicities were also estimated from O3N2 measurements of stacked spectra in bins of stellar mass. The stacked measurements from the complete samples suggest a  decreased oxygen abundance for merging systems compared to isolated systems at a given mass.
}
\label{fig:metal_plots}
\end{figure*}

We now analyze the relationship between stellar mass and metallicity for merging and non-merging galaxies at $2.0\leq z \leq2.7$. We used two different oxygen abundance indicators, N2 and O3N2 \citep{pettini2004}, to check whether our results depended on the indicator adopted. Figure~\ref{fig:metal_plots} displays the distributions of $12+\log(\mbox{O/H})$ and $M_*$ for merging and non-merging galaxies. The left and right panels show, respectively, the MZRs using both our metallicity indicators. 

As in Section~\ref{sec:sfrm}, we performed a Bayesian linear regression to our merging and non-merging samples to find a quantitative relationship between $12+\log(\mbox{O/H})$ and $M_*$. We restrict the stellar mass to $9.0 \leq \log(\mbox{M}_*/\mbox{M}_{\odot}) \leq 11.5$ and require an N2 or O3N2 detection (for individual points) with a S/N ratio $\geq$ 3. Upper limits on $12+\log(\mbox{O/H})$ are shown in Figure~\ref{fig:metal_plots}, but not included in the best-fit regression. The errors associated with the intercept and slope were estimated as in Section~\ref{sec:sfrm}. For merging and non-merging galaxies, we find for N2-based metallicities:

\begin{multline}
    \mathrm{12+log(O/H)_{N2, merger}}= (6.93 \pm 0.43) + \\
    (0.147 \pm 0.042) \times \mathrm{log(M_*/M_{\odot})_{merger}}
\end{multline}

\begin{multline}
    \mathrm{12+log(O/H)_{N2, non-merger}}= (6.48 \pm 0.24) + \\
    (0.196 \pm 0.024) \times \mathrm{log(M_*/M_{\odot})_{non-merger}}
\end{multline}

The linear fit suggests depressed metallicity for merging galaxies compared to non-merging galaxies for a given stellar mass at high redshift. The slope is comparable for both distributions while there is an offset in the intercepts. To further quantify the depression in metallicity, we estimated the average difference between
the best-fit lines for merging and non-merging galaxies over the interval $9.0 \leq \log(M_*/\mbox{M}_{\odot}) \leq 11.5$. Based on the best-fit relations in equations (7) and (8), we find an average of 0.05 dex lower metallicity at fixed stellar mass for merging compared to non-merging systems, when metallicity is estimated with the N2 indicator.

For merging and non-merging galaxies, we find for O3N2-based metallicities:
\begin{multline}
    \mathrm{12+log(O/H)_{O3N2, merger}}= (7.10 \pm 0.45) +\\
    (0.120 \pm 0.045) \times \mathrm{log(M_*/M_{\odot})_{merger}}
\end{multline}

\begin{multline}
    \mathrm{12+log(O/H)_{O3N2, non-merger}}= (6.29 \pm 0.26) + \\
    (0.203 \pm 0.026) \times \mathrm{log(M_*/M_{\odot})_{non-merger}} 
\end{multline}

As with the N2 indicator, the linear fit to the distribution of stellar masses and metallicities based on the O3N2 indicator implies a depressed metallicity for merging systems. The slopes of linear fits to merging and non-merging systems are in agreement, while there is a distinction between the values of the intercepts. 
Although the offset in the intercepts for O3N2-based metallicities is not as prominent as when metallicity is estimated via N2,
both oxygen abundance indicators, N2 and O3N2, imply a lower average metallicity for merging relative to non-merging systems of similar mass and at high redshift.
As in the case of N2, we estimated the average difference between the best-fit lines for merging and non-merging galaxies over the interval $9.0 \leq \log(M_*/\mbox{M}_{\odot}) \leq 11.5$. Accordingly, we find 0.04 dex lower metallicity on average at fixed stellar mass for merging compared to non-merging systems with respect to the O3N2 indicator.

To evaluate the relations in mass and metallicity including information from galaxies with individual limits on metallicities, we constructed median composite spectra in three bins of stellar mass for both merging and non-merging systems using the stacking method presented in \citet{sanders2018}. The results based on stacked spectra are important for gauging if any bias is introduced by only fitting regression relations to galaxies with detections in all of the relevant emission lines. For both the N2 and O3N2 indicators, the lowest-mass bins fall below their respective regression lines when factoring in limits as well as detections, which reflects the bias of only fitting detections. For the majority of stellar mass bins, the metallicities for mergers based on both N2 and O3N2 measurements from composite spectra fall below the corresponding metallicities for non-mergers. We calculate the average offset in metallicity  between the binned points for mergers and non-mergers, based on both N2 and O3N2. We find 
$\langle\Delta(12+\log(\mbox{O/H}))_{\rm N2}\rangle=-0.044\pm 0.025$ and $\langle\Delta(12+\log(\mbox{O/H}))_{\rm O3N2}\rangle=-0.038\pm 0.015$. These offsets are consistent with the results for the linear regression fits to the individual detections of each distribution, and further suggest at the $\sim 2\sigma$ level that metallicity is depressed for merging systems compared to non-merging systems. 

\section{Discussion}
\label{sec:discussion}

\subsection{Comparisons with Other Observational Work}
\label{sec:compare_observations}

Our results represent an extension of the initial analysis of high-redshift mergers in the MOSDEF survey performed by \cite{wilson2019}. We analyze a larger sample of galaxies to explore the relationship between SFR, metallicity, and stellar mass in interacting and isolated galaxies at high redshift. We also use a complementary technique for defining merging systems, based on the CANDELS morphology catalog \citep{kartaltepe2015}. In \cite{wilson2019}, interacting galaxies are identified as {\it spectroscopic} pairs. This selection required at least two distinct objects to be measured within the same spectroscopic slit, with a velocity separation of less than 500~km~s$^{-1}$. \cite{wilson2019} found that merging galaxy pairs do not have elevated SFR or diluted metallicity compared to isolated systems at a given mass. Although significantly larger samples will be required to establish our results with higher significance, we did find such elevations in SFR and depressions in metallicity for mergers relative to non-mergers. The differences in our results relative to those of \cite{wilson2019} may reflect  different techniques of identifying merging systems, and also, in the case of SFR, the fact that we used SFR(H$\alpha$) as our primary SFR indicator. When using SFR(SED), the primary SFR indicator featured in \cite{wilson2019}, we did not find as clear a difference between mergers and non-mergers.

\cite{silva2018} used a peak-finding algorithm to identify merging systems at $0.3 < z <2.5$ in the
CANDELS/3D-HST catalog, with separations of 3 to 15 kpc 
and mass ratios closer or equal to 1:4 (i.e., major mergers). In this work, SFRs are estimated from the combination of rest-UV and mid-IR (i.e., {\it Spitzer}/MIPS 24$\mu$m) luminosities, when IR luminosities were available, and from SED fitting otherwise. Of the selected merging systems at $\log(M_*/M_{\odot}) \geq 10$, only $\sim 12$\% are classified as ``star-bursting," with a deviation in SFR above the star-forming main sequence of $\geq$ 0.5 dex.
\citet{silva2018} also explore the dependence of SFR enhancement on the properties of the merging galaxies, specifically finding larger SFR enhancements in lower-mass ($\log(M_*/M_{\odot}) <10$) galaxies. Overall, however, \citet{silva2018} found no significant difference in the star-forming properties of their merging and non-merging systems, suggesting that these pre-coalescent mergers have yet to reach the maximum enhancement in SFR.

In related work, \citet{cibinel2019} assembled merger samples at $0.2\leq z \leq 2$ defined as either close pairs or else ``morphological mergers," which satisfy morphological merger criteria in the space of the non-parametric Asymmetry and $M_{20}$ statistics. Considered together with non-merger control samples defined over the same redshift range, \citet{cibinel2019} find that the merger sample is offset towards higher SFR at fixed stellar mass. Using a complementary method for describing the differential properties of mergers and non-mergers, \citet{cibinel2019} find that the the merger fraction above the star-forming main sequence (including both types of merger) is $\geq 70$\%. Furthermore, the majority of galaxies falling within the ``starburst" regime at $\Delta \log(\mbox{SFR})=0.6$~dex above the star-forming main sequence are morphologically identified major mergers, whereas the mergers flagged as close pairs scatter symmetrically around the star-forming main sequence. The distinct behavior of close pairs and morphologically-identified mergers in the space of SFR vs. $M_*$ mirrors the fact that \citet{wilson2019} found no evidence for SFR enhancement in the MOSDEF close pairs sample, while we do observe an offset towards higher SFR at fixed $M_*$ for our morphologically-identified sample.

The key result presented in this work is that morphologically-identified mergers in the MOSDEF survey show evidence for enhanced SFR 
and depressed gas-phase metallicity at fixed stellar mass.
Such offsets between merging and non-merging systems
were previously observed in the local universe -- even
when mergers were identified in an earlier stage,
as close pairs \citep[e.g.,][]{ellison2008,scudder2012}.
A coordinated enhancement in SFR and depression in gas-phase metallicity is commonly understood within the framework of the ``fundamental metallicity relation" (FMR) \citep{mannucci2010}. Specifically, at fixed stellar mass, galaxies with higher SFR are offset towards lower metallicity, since the gas accretion that leads to enhanced SFR also dilutes the gas-phase metallicity. The temporal coordination of anti-correlated deviations in SFR and metallicity have also been demonstrated in hydrodynamical simulations of galaxy formation \citep{torrey2018}. 

Intriguingly, however, \citet{bustamante2020} demonstrated that both merging pairs with separation less than 110~kpc and coalesced mergers in SDSS are offset from the FMR towards lower metallicity at fixed stellar mass and SFR, suggesting more more extreme dilution in metallicity in merging systems.  This difference is explained by \citet{bustamante2020} in terms of recently triggered and extremely strong gas inflows in merging systems, whose effects may not be properly captured by the FMR. In future work, based on significantly larger samples with greater statistical significance, it will be vital to quantify the deviations in SFR and metallicity at fixed mass in morphological mergers at high redshift, compared with the expectations from the high-redshift FMR. As part of this analysis, we require a robust measurement of the $z\sim2$ FMR, which is still being established \citep{sanders2018,sanders2020}.

\subsection{Expectations from Simulations}
\label{sec:compare_simulations}
Preliminary studies of the properties of mergers at high redshift, both here and in \citet{cibinel2019}, \citet{silva2018}, and \citet{wilson2019}, suggest that enhancements in SFR and diluted metallicity are more pronounced in morphologically-selected merger samples as opposed to distinct pairs. Such morphologically-selected merger samples -- which include both systems with double or multiple distinct components in close proximity as well as single systems showing signs of structural disturbance -- contain a larger fraction of post-coalescence systems. These results are qualitatively consistent with trends found in the simulations of \citet{fensch2017}. In these simulations, merging $z=2$ galaxies are represented with high (60\%) gas fractions and show only gradual increases in SFR, and only during coalescence. This pattern stands in contrast to what is observed in the low-gas-fraction (10\%) simulations representing local galaxy mergers, where the SFR is boosted more significantly and both prior to and during coalescence. \citet{fensch2017} explain the properties of simulated high-redshift mergers in terms of the nature of gas infall, gas content, and ISM turbulence at high redshift. These simulations do not include cosmological accretion or the globally evolving context of redshift evolution, which are natural components of cosmological simulations such as the IllustrisTNG \citet{pillepich2018} and EAGLE \citet{schaye2015} simulations. 

To date, there have been detailed investigations of the properties of merging galaxies at $z=0-1$ in the IllustrisTNG simulation. \citet{patton2020} found an enhancement in SFR at fixed mass in simulated merging systems in the IllustrisTNG simulations, which increases with decreasing pair separation, and agrees well with the observed SFR enhancements in low-redshift SDSS galaxies. The trend of increasing SFR at smaller scales becomes less pronounced as redshift increases from 0 to 1. \citet{hani2020} conducted a similar analysis of post-merger systems at $z=0-1$ in IllustrisTNG, finding a significant enhancement in their SFR relative to a sample of non-interacting controls, and no significant redshift evolution. Based on the Auriga high-resolution cosmological simulations of galaxy formation, \citet{bustamante2018} found not only enhanced star formation, but also dilution in the metallicity of simulated merging systems at $z=0-1.5$. As is observed among merging SDSS galaxies, the observed magnitude of the dilution in metallicity exceeds predictions from the FMR.

With the advent of new observations of the star-forming and chemical properties of merging pairs and morphologically-identified mergers at $z\sim 2$, it will be important to extend the analysis of merging systems in {\it cosmological} simulations \citep[e.g.,][]{patton2020} to higher redshift. The trend of declining enhancement in SFR at higher redshift over $z=0-1$ in the IllustrisTNG simulation \citep{patton2020} should be extended back to $z\sim 2$. Such an analysis can be used to test the claims about the decreased effect of mergers on SFR at high redshift, based on the non-cosmological simulations of \citet{fensch2017}.

\subsection{Future Work}
\label{sec:future}
Our analysis continues to characterize properties of merging galaxies at high redshift. We looked at a collection of 250 galaxies, 55 of which were confidently classified as merging systems on the basis of morphology. 
Although our sample of mergers is larger than the one presented in  \cite{wilson2019}, it is still too small to draw definitive conclusions about the relationships between merging and non-merging systems at high redshift. Looking ahead, we must assemble samples on the order of $>1000$ merging pairs as in \cite{ellison2008} and \cite{scudder2012}. Both studies divided merging systems into bins of mass ratio and radial separation, exploring the characteristics of mergers based on physical properties, and contained large enough numbers of galaxies that the mean properties of mergers and non-mergers were established with small error bars. In order to trace the properties of mergers in bins of both mass and separation with high confidence, we require a significant boost in statistics. 

\section*{Acknowledgements}
We thank the referee for a constructive and thorough report, which significantly improved the paper. We acknowledge support from NSF AAG grants AST-1312780, 1312547, 1312764, and 1313171, archival grant AR-13907 provided by NASA through the Space Telescope Science Institute, and grant NNX16AF54G from the NASA ADAP program. We also acknowledge a NASA contract supporting the ``WFIRST Extragalactic Potential Observations (EXPO) Science Investigation Team" (15-WFIRST15-0004), administered by GSFC. We also thank Jeyhan Kartaltepe for sharing the CANDELS morphological merger catalog.
We wish to extend special thanks to those of Hawaiian ancestry on whose sacred mountain we are privileged to be guests. Without their generous hospitality, most
of the observations presented herein would not have been possible

\section*{Data Availability}
The data underlying this article will be shared on reasonable request to the corresponding author.

\bibliographystyle{mnras}
\bibliography{ms}

\end{document}